\documentclass[referee,useAMS,usenatbib]{mn2e}

\title[Observational evidence for disk-SMBBHs interaction at pc 
  scale]{X-shaped radio galaxies as observational evidence for the
  interaction of supermassive binary black holes and accretion
  disk at pc scale} 
\author[F.K. Liu]{F. K. Liu\thanks{E-mail:
    fkliu@bac.pku.cn}\\
Astronomy Department, Peking University, 100871 Beijing, China}

\begin{document}

\date{Accepted ***. Received ***; in original form ***}

\pagerange{\pageref{firstpage}--\pageref{lastpage}} \pubyear{2002}

\maketitle

\label{firstpage}

\begin{abstract}
  In the hierarchical galaxy formation model, today's galaxies are the
product of frequent galaxy merging, triggering the activity of active
galactic nuclei and forming a supermassive black hole binary. A binary
may become stalling at pc-scale and is expected to be detected in
nearby normal galaxies, which is inconsistent with observations.  
In this paper, we investigate the interaction of the supermassive 
binary black holes (SMBBHs) and an accretion disk and show that the 
stalling can be avoided due to the interaction and a rapid coalescence
of SMMBHs can be reached. A binary formed during galaxy merging
within Hubble is most likely inclined with a random inclination angle
and twists the accretion disk, aligning the inner part of the disk
with the orbital plane on a time scale $\sim 10^3 \, {\rm yr}$. The
twisted inner disk subsequently realigns the rotating central 
supermassive black hole on a time scale $\la 10^5 \, {\rm yr}$
due to the Bardeen-Peterson effect. It is shown that the detected
X-shaped structure in some FRII radio galaxies may be due to the
interaction-realignment of binary and accretion disk occurred within
the pc scale of the galaxy center. The configuration is consistent
very well with the observations of X-shaped radio sources. X-shaped
radio feature form only in FRII radio sources due to the strong
interaction between the binary and a standard disk, while the 
absence of X-shaped FRI radio galaxies is due to that the interaction
between the binary and the radiatively inefficient accretion flow in
FRI radio sources is negligible. The detection rate, $\lambda_{\rm X}
\sim 7 \%$, of X-shaped structure in a sample of low luminous FRII 
radio galaxies implies that X-shaped feature forms in nearly all FRII
radio sources of an average lifetime $t_{\rm life} \sim 10^8 \, {\rm
yr}$. This is consistent with the estimates of net lifetime of QSO and
radio galaxies and with the picture that the activity of active
galactic nuclei is triggered by galaxy merging. As the jet orients
vertically 
to the accretion disk which is supposed to be aligned with galactic
plane of host galaxy, the old wings in X-shaped radio sources are
expected to be aligned with the minor axis of host galaxy while the
orientation of the active jets distributes randomly. It is suggested
by the model that the binary would keep misaligned with the outer disk
for most of the disk viscous time or the life time of FRII radio
galaxies and the orientation of jet in most FRII radio galaxies
distributes randomly. As the binary-disk interaction in FRI radio
galaxies is negligible or a source evolves from FRII- to FRI-type
after the binary becomes aligned with the outer disk, the jets in 
most FRI radio galaxies is expected to be vertical to the accretion
disk and thus the major axis of host galaxy. We discuss the 
relationship of X-shaped and double-double radio galaxies
(DDRGs) and suggest that all X-shaped radio sources would evolve into
DDRGs after the coalescence of the SMBBHs and that most radio sources
evolve from FRII- to FRI-type after an interruption of jet formation,
implying that the average size of FRI radio sources is smaller than
that of FRII radio galaxies. The model is applied to two X-shaped 
radio sources 4C+01.30 and 3C293 and one DDRG source J0116-473 with a
bar-like feature and show that the SMBBHs in the three objects are
minor with mass ratio $q \sim 0.1 - 0.3$. 
\end{abstract}


\section[]{Introduction}
\label{sec:intr}

Recent observations show that all galaxies with bulges harbor a 
supermassive black hole (SMBH) of mass tightly correlating with 
both the mass and the velocity dispersions of the bulge 
\citep{ferrarese00,gebhardt00,magorrian98,mclure02,tremaine02}. This
implies that one event is responsible for both the formations of the
bulges and the central SMBH.
 
In hierarchical galaxy formation models, present-day galaxies are the
product of successive minor mergers \citep{kauffmann00,haehnelt00,
menou01}, triggering the star-bursts and the activity of active
galactic nuclei (AGNs) \citep{wilson95}. The configuration is 
supported by the observations of compact steep-spectrum objects 
(CSSs) and gighertz-peaked sources (GPSs) which are most likely 
infant AGNs of age $t \la 10^5 \, {\rm yr}$ and show distortion of
host galaxy, double nuclei, galaxy-interaction or close companion
among about fifty per-cent of them \citep{odea98}. In galaxy mergers,
the two SMBHs at the galaxy centers become bound due to dynamical 
friction at a separation $a$ of two SMBHs, $a \sim 1 - 10 \, {\rm
pc}$, and become hard at a separation $a_{\rm h} \sim 0.01 - 1 \, 
{\rm pc}$ when the loss of the orbital angular momentum is dominated
by three-body interactions between SMBBHs and the stars passing by and
a loss cone forms \citep{begelman80,quinlan97,yu02,makino03}. The
decay time scale of the binary orbit is then propotional to the
relaxation timescale of the parent galaxy longer than the cosmological
time and the SMBBHs become stalling at $a \sim a_{\rm h}$. Therefore,
it is expected to detect SMBBHs in many nearby normal galaxies or AGNs.
However, efforts to detect long-lived SMBBHs in normal galaxies are
failed \citep{haehnelt02} and SMBBHs may be detected only in a handful
AGNs, e.g. OJ287 \citep{sillanpaa88,liu02b}, ON231 \citep{liu95b}, 
PKS1510-089 \citep{xie02}, MKN421 \citep{liu97}, NGC6260 
\citep{komossa03}, and 3C66B \citep{sudou03} (for a review of
observations of SMMBHs, see \citet{komossa03a}). \citet{liu03} suggest
that SMMBHs in FRII radio galaxies merger during FRII-active phase,
leading to the formation of DDRGs. Many stellar dynamical mechanisms
have been invoked to extract the angular momentum in literature (for a
review, see \citet{milosavljevic02}) and found to be inefficient, 
including black hole wondering \citep{chatter03}, stellar slingshot
effects and re-filling of the loss cone \citep{zier01}, and the Kozai
mechanism \citep{blaes02}. This is the so-called {\it final parsec
problem} \citep{milosavljevic02}. 

Gas in merging galaxies is driven to the center on a time scale 
$t_{\rm g} \approx 10^8 \, {\rm yr}$ and triggers the star-bursts 
and AGN activity \citep{gaskell85,hernquist95,barnes96,barnes02},
forming an accretion disk around central SMBH and of size as large
as $r_{\rm d} \sim 10\, {\rm pc}$ \citep{jones00} and in
general $r_{\rm d} \sim 10^4 r_{\rm G} \sim 0.01 - 1 \, {\rm pc}$ 
\citep{collin01}, where $r_{\rm G}$ is the Schwarzschild radius of
central black hole. Therefore, it is expected that the secondary 
interacts with the accretion disk whenever a binary becomes hard.
The interaction of an accretion disk and a binary is intensively 
investigated in literature (e.g. \citet{lin86,artymowicz94,ivanov98,
ivanov99,gould00,armitage02,narayan00,liu03}). If an accretion disk 
is standard, initially inclined SMBBHs warp, twist and align the
accretion disk within some radius $r_{\rm al} > a$ \citep{ivanov99}.
Then, the orbital plane slowly becomes aligned with the outer disk of
$r> r_{\rm al}$ due to exchange of angular momentum with accreting gas
on a time scale depending on the total disk mass. When the orbital
plane completely becomes coplanar with the outer accretion disk, the
secondary opens a gap in the disk and losses its angular momentum via
viscosity torque and in-spiraling on a viscous time scale 
\citep{lin86,armitage02}. When the loss of orbit angular momentum due
to gravitational wave radiation becomes dominated at $a \la 10^2
r_{\rm G}$, the secondary removes mass of the inner disk and mergers
into the primary, leading to the interruption of jet formation 
\citep{liu03},  

\Citet{liu03} identify DDRGs with objects in which the coalescence of 
SMBBHs and the removal of inner disk occurred. Although the 
observations of DDRGs are consistent the scenario, \citet{merritt02}
suggest that binary coalescence leads to a spin-flip of central SMBH
and forms X-shaped radio galaxies. So, the important question is
which picture is correct or, if both are correct, what the
relationship between DDRGs and X-shaped radio sources is. As there 
are several difficulties with the Merritt \& Ekers' configuration
(see the discussions in Sec.~\ref{sec:others}), we show in this paper 
that the mechanism to form the X-shaped feature detected in some FRII
radio galaxies may be the interaction between the binary and an 
standard accretion disk at pc scale and that X-shaped feature and
double-double lobes forms in different evolution phases of SMBBHs in 
FRII radio galaxies. 

We investigate the interaction of SMBBHs and an accretion disk at
pc scale and the reorientation of spin axis of rotating central
SMBH and compare this configuration with the observations of X-shaped
radio galaxies, which are consistent with each other. In this
scenario, the FRII character of X-shaped radio galaxies is due to that
the accretion disk is an standard $\alpha$-disk in FRII radio sources,
while in FRI radio galaxies the accretion flow is radiatively
inefficient ADAFs and the interaction of an ADAF and a binary is
negligible. Basing on this scenario, we predicate that the
orientations of active radio jets distribute randomly in FRII radio
galaxies and are preferentially vertical to the major axis of host
galaxies in FRI radio sources. The model also suggests that radio jets
are  aligned with the minor axis of host galaxies in DDRGs. 

In the paper, we discuss the initial conditions of the binary-disk
system basing the Bardeen-Peterson effect and specify the SMBBH system
in Sec.~(\ref{sec:model}). The interaction of accretion disk and
inclined SMBBHs is investigated in detail in Sec.~\ref{sec:dbhit}. 
In this section, we pay our special attention to the reorientations
of the binary orbit and rotating central SMBH. In 
Sec.~\ref{sec:formation}, we discuss the connection between X-shaped
radio galaxies and the objects in which the reorientation of spin axis
of rotating central SMBH due to disk-binary interaction occurred. All
the observations of X-shaped radio sources are discussed basing
on the scenario. We also discuss the distribution of orientations of
wings and active jets in X-shaped and normal radio galaxies in this
section. The important question what is the relationship between DDRGs
and X-shaped radio sources is addressed in Sec.~\ref{sec:ddrg}. Our
discussions and conclusions are presented in Sec.~\ref{sec:dis}.

\section[]{Final parsec problem of the binary evolution and the
  formation of disk-SMBBH system}
\label{sec:model}

\subsection[]{Bardeen-Peterson effect and alignment of rotating SMBBH 
  and inclined accretion disk}
\label{sec:bp}

When the part of cold gas with low angular momentum loses its angular
momentum due to viscosity and flows in-wards toward the central
SMBH, an accretion disk and relativistic jets form \citep{shlosman90,
barnes02}. As the formed accretion disk keeps the angular momentum of 
gas, it is nearly aligned with the galactic gas disk, if no strong
torque is exerted on it. Since all young AGNs with newly born 
relativistic jets of age have FRII radio morphologies 
\citep{odea98,murgia02,perucho02}, the accretion disk 
has an accretion rate $\dot{m} \equiv 
{\dot{M} /\dot{M}_{Edd}} = {L / L_{\rm Edd}} \ga \dot{m}_{\rm FR}
= 6 \times 10^{-3}$ \citep{ghisellini01,cavaliere02,maraschi03} and 
is a standard $\alpha$-disk if $\dot{m} \la
1$ \citep{shakura73} or slim disk for $\dot{m} \ga 1$
\citep{abramowicz88}. Here, $\dot{M}_{Edd} = L_{Edd} / 
\epsilon c^2 = 2.30 M_8 \, {\rm (M_\odot \, yr^{-1})}$ for $\epsilon =
0.1$ is the Eddington accretion rate (note the different definitions
of Eddington accretion rate), $\epsilon = 0.1 \epsilon_{-1} = L
/\dot{M} c^2$ is the conversion rate of accretion mass to energy and
$M = M_8 \times 10^8 M_\odot$ is the mass of central SMBH.  

For a gas-pressure dominated standard $\alpha$-disk, the ratio
$\delta$ of the half thickness $H$ of accretion disk and radius $r$ is 
\begin{eqnarray}
\delta & \equiv & {H \over r} \nonumber \\
       &\simeq& 2.8 \times 10^{-3} \alpha_{-2}^{-1/10} 
         \dot{m}_{-1}^{1/5} M_8^{-1/10} x_4^{1/20} ,
\label{eq:delta1}
\end{eqnarray}
where $\alpha_{-2} = \alpha /0.01$, $\dot{m}_{-1} = \dot{m}/0.1$, $x_4
= r/10^4  r_{\rm G} $ and $r_{\rm G} = 2 G M /c^2 = 2.97
\times 10^{13} M_8 \, {\rm cm}$ is the Schwarzschild radius. 
Such a gas-pressure dominated disk has a surface mass density
\begin{equation}
  \Sigma \simeq 4.0\times 10^5 \alpha_{-2}^{-4/5} \dot{m}_{-1}^{3/5}
  M_8^{1/5} x_4^{-3/5} \, {\rm g \; cm^{-2}} 
\end{equation}
and a total mass within the disk radius $r_{\rm d}$
\begin{equation}
  M_{\rm d} = { 10 \pi \over 7} \Sigma r_{\rm d}^2 \simeq 7.92\times
  10^7 \alpha_{-2}^{-4/5} \dot{m}_{-1}^{3/5} 
  M_8^{11/5} x_4^{7/5} \,  {\rm M_\odot} .
\label{eq:tdm}
\end{equation}
The disk radius $r_{\rm d}$ is empirically \citep{collin01}
\begin{eqnarray}
  r_{\rm d} & \approx & 2\times 10^4 M_7^{-0.46} r_{\rm G} \nonumber\\
            & \simeq & 7 \times 10^3 M_8^{-0.46} r_{\rm G} .
\label{eq:dsize}
\end{eqnarray}
However, accretion disk is no longer gas-pressure dominated for $r \la
10^2 r_{\rm G}$ and the disc opening angle is \citep{collin90}
\begin{equation}
\delta  \simeq 9.9\times 10^{-3} \alpha_{-2}^{-1/10} 
         \left({L \over 0.1 L_{\rm Edd}}\right)^{1/5} M_8^{-1/10}
	\epsilon_{-1}^{-1/5} x^{1/20} .
\label{eq:delta2}
\end{equation}

For an accretion disk in AGNs of a typical value $\alpha \sim
 0.03$, $ \delta < \alpha \ll 1$. If the rotating 
central SMBH is misaligned with the accretion disk, the innermost part of
the accretion disk becomes aligned with SMBH spin direction due to the
Bardeen-Peterson effect \citep{bardeen75} out to a disk radius $r_{\rm
BP} \gg r_{\rm G}$. When the rotating SMBH exerts a torque on
the accretion disk and aligns the innermost part of the disk with its 
spin, the same torque tends to align the spin of the SMBHs with the
accretion disk \citep{rees78,scheuer96}, depending on the transfer of 
warps in radial direction. For an accretion disk
with $\alpha > \delta$ in AGNs, warps transfer in a
diffusive way. \citet{papaloizou83} show that taking into
consider of the internal hydrodynamics of the disk the usual 
azimuthal viscosity $\nu$ and the viscosity $\nu_2$ in the vertical
direction are different. The inward advection of angular momentum
via $\nu$ is rather accurately canceled by the outward viscous
transport of angular momentum due to $\nu$, while the radial 
pressure gradients due to the warp set up radial flows, whose natural
period resonates with the period of the applied force and therefore
reaches large amplitude. Finally, the effective vertical viscosity 
is approximately given by $\nu_2 = \nu /2 \alpha^2$
\citep{papaloizou83,kumar85}, which is valid even for a significant 
warp \citep{ogilvie99}. Thus, the transfer time-scale of warp in the 
accretion disk is 
\begin{equation}
  t_{\rm wp} \simeq {2 r^2 \over 3 \nu_2} = 2 \alpha^2 t_{\rm v} =
  2\times 10^{-4} \alpha_{-2}^2 t_{\rm v} ,
\label{eq:wq}
\end{equation}
where 
\begin{equation}
  t_{\rm v} = {r \over v_{\rm r}} = {2 r^2 \over 3 \nu} = {2 \over 3}
  \delta^{-2} \alpha^{-1} \Omega_{\rm K}^{-1} ,
\label{eq:vist}
\end{equation}
where $\Omega_{\rm K}$ is the Keplerian angular velocity at radius $r$
and $v_{\rm r} = 3\nu /2 r$ is the flow velocity in radial direction.

Considering the difference of $\nu_2$ and $\nu$ and using the
accretion disk models for AGNs computed by \citet{collin90}, 
\citet{natarajan98} show that the Bardeen-Peterson radius out to which the
inner accretion disk of $L/L_{\rm Edd} = \dot{m} > \dot{m}_{\rm cr}$
is aligned with the rotating SMBH is 
\begin{equation}
  r_{\rm BP} = 22 a_*^{5/8} \alpha_{-2}^{3/4} \left({L \over 0.1
  L_{\rm Edd}}\right)^{-1/4} M_8^{1/8} \epsilon_{-1}^{1/4} 
\end{equation}
and the spin axis of a rotating SMBH becomes aligned with the
accretion disk due to the Bardeen-Peterson effect on a time scale 
\begin{eqnarray}
t_{\rm al1}  & = & 3.6 \times 10^4 \, {\rm yr}\, a_*^{11/16}
      \alpha_{-2}^{13/8} \times \nonumber \\
      & & \left({L \over 0.1 L_{\rm Edd}}\right)^{-7/8}
      M_8^{-1/16} \epsilon_{-1}^{7/8}  ,
\label{eq:at1}
\end{eqnarray}
where $a_*$ is the dimensionless spin angular momentum of the primary and
$L$ is the luminosity of AGNs. Eq.~\ref{eq:at1} shows that the
realignment time-scale $t_{\rm al1}$ is nearly independent of the SMBH
mass but  sensitive to the parameter $\alpha$. For an AGN with a 
moderately rotating central black hole $a_* \simeq 0.7$, and typical
parameters $\alpha_{-2} = 3$, $\epsilon_{-1} = 2$ and $L \sim 0.3 L_{\rm
  Edd}$, the realignment time scale is $t_{\rm al1}
\simeq 1.1\times 10^5 \, {\rm yr}$. If we take a typical advance speed
of radio lobes $v_{\rm j} \simeq 0.3 c$ for a young AGN
\citep{owsianik99} (we assumed an accelerating cosmology with
$\Omega_\Lambda = 0.7$, $\Omega_B = 0.3$ and $H_0 = 70 \, {\rm km \;
s^{-1} \; Mpc^{-1}}$ throughout this paper except 
mention), the realignment takes place when the source has a largest
linear size $l_{\rm m} \la t_{\rm al1} v_{\rm j} \simeq 10 \, {\rm
Kpc}$. An AGN is detected with a GPS source if $l_{\rm m} < 1 \, {\rm
Kpc}$ or a CSS source if $l_{\rm m} \la
20 \, {\rm Kpc}$ (for $H_0 = 100 \, {\rm km \; s^{-1} \; Mpc^{-1}}$,
$q_0 = 0.5$) \citep{odea98}. Therefore, dramatical distortion may 
be detected in some radio jets of CSSs and GPSs. When the realignment
finishes, lobes randomly orient and jets close to the central nuclei
are nearly vertical to the accretion disk.

\subsection[]{The hardening of SMBBHs and the interaction of the 
  secondary and an accretion disk}

When the gas in merger galaxies with low angular momentum is driven
into center and  triggers the activity of AGNs, the two 
SMBHs at the two galaxy centers lose the angular momentum 
due to dynamic friction and become bound at a
separation of the two black hole $ a \sim 1 - 10 \, {\rm pc}$
on a time scale \citep{merritt00}
\begin{equation}
t_{\rm bd} \simeq {r_{\rm e} \over 0.30} {\sigma^2 \over \sigma_{\rm
    g}^3} ,
\label{eq:bd0}
\end{equation}
where $\sigma$ and $\sigma_{\rm g}$ are, respectively, the
one-dimensional velocity dispersions of the larger (primary) and
smaller (secondary) galaxies, and $r_{\rm e} \simeq 2.6 \, {\rm Kpc} \,
(\sigma/200 \, {\rm Km\; s^{-1}})^3$ is the effective radius of the
larger galaxy. Relating the one-dimensional velocity dispersions
$\sigma$ to the mass of the central SMBH with the empirical tight
relation \citep{tremaine02} \begin{equation}
  \log(M / M_\odot) = 8.13 + 4.02 \log(\sigma / 200 \, {\rm km \;
    s^{-1}}) 
\label{eq:mass}
\end{equation}
and from Eq.~(\ref{eq:bd0}), we have
\begin{equation}
  t_{\rm bd} \simeq 2.0 \times 10^8 \, {\rm yr} \, M_8^{1/2.01}
  q_{-1}^{-3/4.02} ,
\label{eq:tbd}
\end{equation}
where $q \equiv m/M = 0.1 q_{-1}$.  Eq.~\ref{eq:tbd} implies that 
$t_{\rm bd} \sim t_{\rm g}$ and is consistent with the observations
of host galaxies of young AGNs that about 50 percent of the host 
galaxies contain double nuclei, interaction of galaxies or 
significant morphological distortions due to galaxy merging
\citep{odea98}.  

From Eq.~\ref{eq:tbd}, $t_{\rm bd}$ is larger than the Hubble time 
$t_{\rm Hubble}$ when $q < 5 \times 10^{-4} M_8^{2/3} (t_{\rm bd}
/10^{10} \, {\rm yr})^{-4/3}$, which is consistent with the numerical
galaxy dynamical calculations \citep{yu02}. As we are interested only 
in those SMBBHs formed within Hubble time, we have 
\begin{equation}
  q > q_{\rm cr} \simeq 5\times
  10^{-4} M_8^{2/3} \left({t_{\rm bd} \over 10^{10} \, {\rm
  yr}}\right)^{-4/3}  .
\label{eq:qcr}
\end{equation}
Bound SMBBHs become hard at a separation of the two SMBHs
\citep{quinlan96,yu02} 
\begin{eqnarray}
  a_{\rm h} & = & {G mM \over 4 \sigma^2 (m+M)} \nonumber \\
            & = & 6.5 \times 10^3 M_8^{-1/2.01} \left({q \over 
            0.02}\right) \left(1+ q\right)^{-1} r_{\rm G} 
\label{eq:ahd}
\end{eqnarray}
on a time scale $\la t_{\rm bd}$ due to dynamic friction. A hard binary
loses orbital angular momentum via stellar dynamic interaction on a
time scale maybe much longer than the Hubble time \citep{begelman80,
quinlan96,yu02} (for more discussion, see Sec.~\ref{sec:intr}). 
However, Eqs.~\ref{eq:dsize} and \ref{eq:ahd} show that $a_{\rm h} \la
r_{\rm d}$ for a minor merger with $q \la 0.3$ and the binary
interacts with the accretion disk soon after it becomes hard. As the
interaction between the SMBBHs and a standard accretion disk is very
efficient in hardening SMBBHs, the {\it final parsec problem} can be
avoided. We investigate in detail the interaction between the SMBBHs
and an accretion disk in following sections. 

When the orbital radius $a$ is very small, the loss of angular
momentum of the SMBBHs due to gravitational radiation becomes
important. When $a$ is smaller than a critical radius $a_{\rm cr}$,
the loss of angular momentum due to gravitational radiation becomes
dominated with respect to binary-disk interaction. \citet{liu03}
show 
\begin{equation}
    a_{\rm cr} = {1 \over 2} \left({128 \over 15}\right)^{2/5} 
               \delta^{-4/5} \alpha^{-2/5} q^{2/5} \left(1 + 
	       q\right)^{1/5} f^{2/5} r_{\rm G} ,
\label{eq:acr}
\end{equation}
where $f$ is a function of the binary eccentricity $e$:
\begin{equation}
f = \left(1 + {73 \over 24} e^2 + {37 \over 96} e^4\right) 
	\left(1 - e^2\right)^{-7/2} .
\label{eq:fec}
\end{equation}
From Eqs.~\ref{eq:ahd}, \ref{eq:acr} and \ref{eq:delta1}, the
interaction between binary and accretion disk is always important 
only if 
\begin{equation}
  q > 6.0 \times 10^{-5} M_8^{29/30} \alpha_{-2}^{-8/15}
  \dot{m}_{-1}^{-4/15} x_2^{-1/15} f^{2/3} ,
\end{equation}
where $x_2 = r/10^2 r_{\rm G}$.  Therefore, the interaction of the
accretion disk and the SMBH plays a very important role in the
evolution of SMBBHs with $q >  5 \times 
10^{-4} M_8^{2/3} (t_{\rm bd} /10^{10} \, {\rm yr})^{-4/3}$.

\section[]{The alignment of accretion disk with orbital
  plane and the reorientation of central rotating SMBH}
\label{sec:dbhit}

\subsection[]{Accretion modes in AGNs}

Although the accretion disk in young AGNs is most likely a standard
thin- or slim-disk, the accretion rate may have decreased 
significantly when the binary becomes hard and interacts with the
accretion disk. When the secondary SMBH enters the accretion disk with
a random inclination angle, it interacts with the disk in two regimes:
direct collisions and long-ranged interaction. The interaction between
a companion star and the accretion disk around a primary has been
investigated intensively in literature (e.g. \citet{ivanov98,ivanov99,
lin86,narayan00}) and the effects of the interaction depends on 
accretion modes and disk total mass.

If the accretion rate $\dot{m}$ of a disk is lower than a critical
rate $\dot{m}_{\rm cr} \sim 10^{-2} - 10^{-1}$, the inner disk or the
whole accretion disk becomes, probably via disk evaporation 
\citep{meyer94,liu95a,meyer00}, radiatively inefficient advection
dominated accretion flows (ADAFs) \citep{narayan94,abramowicz95},
advection dominated inflow and outflow (ADIOs) \citep{blandford99}, or
convection dominated accretion flows (CDAFs) \citep{narayan00a}. 
\citet{narayan95} and \citet{esin97} show that if the fraction of
viscously dissipated energy advected is $f \simeq 1$, the transitional
accretion rate is $\dot{m}_{\rm cr} \sim 1.3 \alpha_{\rm AD}^2$, 
where $\alpha_{\rm AD}$ is the viscous parameter of the radiatively
inefficient accretion flow. Both dynamical investigation \citep{liu02b} and
spectrum fit \citep{esin97,quataert99} show that $\alpha_{\rm AD}$ is
typically between 0.3 and 0.1.
As the critical accretion rate corresponds to the transition
from a radiation dominated regime of $f \simeq 0$ to advection
dominated accretion of $f\simeq 1$, it is expected that $0 < f < 1$
and $\dot{m}_{\rm cr}$ critically depends on the physical process in
the disk. \citet{mahadevan97} shows 
\begin{eqnarray}
\dot{m}_{\rm cr} & = & 7.8 {\left(1 - f\right) \over f}  {\left(1 - 
          \beta\right) \over \beta} \alpha_{\rm AD}^2 c_1^2 {1 \over 
          g(\theta_{\rm e})} \nonumber \\
          & \simeq & 0.28 \alpha_{\rm AD}^2 ,
\label{eq:adcrit}
\end{eqnarray}
where $\beta$ is the ratio of gas pressure to total pressure, $c_1
\simeq 0.5$, and $g(\theta_{\rm e})$ is a function of electron
temperature with $g(\theta_{\rm e}) \sim 7$. In computing
Eq.~(\ref{eq:adcrit}), \citet{mahadevan97} takes $f\sim 0.5$ and
$\beta \sim 0.5$, which is consistent with the recent suggestion
that the transition most likely occur at the region where radiation 
pressure becomes important. If the FR transition of radio galaxies
corresponds to the transition of accretion mode $\dot{m}_{\rm FR} 
\simeq 6\times 10^{-3}$ implies that the
typical viscous parameter in FRI radio galaxies is $\alpha_{\rm AD}
\simeq 0.15$. Accretion disk is believed to be thin \citep{shakura73}
for $1 \ga \dot{m} \ga \dot{m}_{\rm cr}$ and slim \citep{abramowicz88}
when $\dot{m} \ga 1$.


\subsection[]{Interaction between SMBBHs and standard thin disks}
\label{sec:stand}

As the outer part of an accretion disk with $r \ga 10^2 r_{\rm G}$
is most possibly a gas pressure-dominated standard disk, we first 
explore its interaction with SMBBHs. From Eqs.~(\ref{eq:tdm}) and 
(\ref{eq:dsize}), we obtain the ratio $\eta$ of the disk total mass
and the mass of the secondary
\begin{eqnarray}
\eta & \equiv &  {M_{\rm d} \over m} \nonumber \\
     & \approx & 5 q_{-1}^{-1} \alpha_{-2}^{-4/5} \dot{m}_{-1}^{3/5}
     M_8^{0.556} .
\label{eq:mmd}
\end{eqnarray}
For those AGNs or QSOs with super Eddington accretion $\dot{m} \ga 1$,
$\eta$ is always greater than unit for any central BH masses in the
range from $10^6 \, {\rm M_\odot}$ to $10^{10} \, {\rm M_\odot}$
\citep{wu02}. For a minor merger with mass ratio $q < 0.3$ and of a
standard thin disk of accretion rate $\dot{m} > \dot{m}_{\rm cr} \sim
10^{-2} - 10^{-1}$, $M_{\rm d} \ga m$. 

When the secondary enters the accretion disk with a radius $a \simeq
r_{\rm d}$, the disk mass $M_{\rm b}$ contained inside its orbit is
$\sim M_{\rm d}$ and $M_{\rm b} \sim \eta m \ga m $ and the
interaction between the secondary and accretion disk is dominated
by BH-disk direct collisions and the distortion of the accretion disk
due to the long range interaction of the secondary is small
\citep{ivanov98,ivanov99}. \citet{ivanov98} and \citet{vokrouhlicky98}
show that during each direct collision 
an amount of gas of mass $\sim \pi \Sigma r_{\rm a}^2$ is accreted by
the secondary and another amount of approximately the same mass gets
velocities larger than the escape velocity in the SMBBH potential and
leaves the system, where the accretion radius $r_{\rm a}$ of the
secondary is
\begin{equation}
  r_{\rm a} = { 2 G m \over v_{\rm rel}^2} \sim { 2 G m \over v_{\rm
  K}^2} = 2 q r
\end{equation}
and $v_{\rm rel} \sim v_{\rm K}$ is the velocity of the secondary
relative to the disk gas. The outflow rate of the mass due to the
direct collision is 
\begin{equation}
  \dot{M}_{\rm out} \sim 2\times {2 \pi \Sigma r_{\rm a}^2 \over
  t_{\rm orb}} = 2 \Sigma r_{\rm a}^2 \Omega_{\rm K} .
\end{equation}
The disk drag to the secondary is important only when the outflow is
compensated by the radial inflow of mass, i.e. $\dot{M}_{\rm out} <
\dot{M} = 3 \pi \nu \Sigma = 3 \pi \alpha \delta^2 r^2 \Omega_{\rm K} 
\Sigma$, which implies 
\begin{eqnarray}
  q < q_0 & = & \sqrt{3 \pi \over 8} \alpha^{1/2} \delta \nonumber \\
          & \simeq & 3\times 10^{-4} \alpha_{-2}^{2/5} 
          \dot{m}_{-1}^{1/5} M_8^{-1/10} x_4^{1/20} .
\label{eq:qcol}
\end{eqnarray}
Eqs.~(\ref{eq:qcr}) and (\ref{eq:qcol}) suggest that for SMMBHs with
$q> q_{\rm cr}$ formed during galaxy merging within Hubble time 
the depleted region by the secondary cannot be sufficiently refilled
with gas and its surface density is thus much less than the
unperturbed value. The secondary would lose it angular momentum and
migrates inwards on accretion time-scale $\sim t_{\rm acc}$
\citep{ivanov99} 
\begin{equation}
  t_{\rm acc} \equiv { m \over  \dot{M}} = {q \over \dot{m}} t_{\rm s}
  = 4.5\times 10^7 \, {\rm yr} \, q_{-1} \dot{m}_{-1}^{-1} ,
\label{eq:acc}
\end{equation}
where $t_{\rm s} \equiv M/\dot{M}_{\rm Edd} = 4.5\times 10^7 \, {\rm 
yr}$. 

When the secondary migrates toward the mass center of binary and has
a orbital radius $a$ less than a critical radius $a_{\rm m}$ within 
which the disk mass $M_{\rm b}$ is equal to its mass $m$, the 
long-ranged averaged quadrupole component of the binary gravitational
field becomes important and warps the disk first in the vicinity of 
the orbit \citep{lin86,ivanov99,kumar90}. From Eq.~(\ref{eq:tdm}), we
have 
\begin{equation}
  a_{\rm m} = 2.3\times 10^3 q_{-1}^{5/7} \alpha_{-2}^{4/7}
  \dot{m}_{-1}^{-3/7}  M_8^{-6/7} r_{\rm G} 
\label{eq:am}
\end{equation}
and $a_{\rm m} \gg r_{\rm BP}$. 
The twist and warp transfers both inwards and outwards. Numerical
simulation given by\citet{ivanov99} shows that the inner accretion
disk evolves into a quasi-stationary twisted configuration and 
becomes coplanar with the binary orbital plane for any value of
$\alpha$ \citep{ivanov99}. The alignment time scale $t_{\rm al2}$
is the warp transfer time scale $t_{\rm tw}$ and for an accretion 
disk with $\alpha > \delta$ the transfer is radiative with
\begin{equation}
  t_{\rm al2} \simeq 3.0 \times 10^3 \, {\rm yr} \, 
               \alpha_{-2}^{6/5} {\dot m}_{-1}^{-2/5} M_8^{6/5}
               x_3^{7/5} ,
\label{eq:alt2}
\end{equation}
where $x_3 = a /10^3 r_{\rm G}$. To obtain Eq.~(\ref{eq:alt2}), we have
used Eq.~(\ref{eq:wq}), (\ref{eq:vist}) and (\ref{eq:delta1}). For an
accretion disk with $\alpha < \delta$, the transfer is wave-like
\citep{ivanov99} and the time scale $ t_{\rm al2}$ is $\sim a / c_{\rm
s} \simeq \delta^{-1} \Omega_{\rm K}^{-1} \sim 1.4 \, {\rm yr} 
\delta^{-1} x_3^{3/2} M_8$. However, the innermost region of the
aligned disk in the vicinity of the primary SMBH with $r \la r_{\rm
bp}$ is twisted and aligned with the rotating central SMBH due to
the Bardeen-Peterson effect. The rotating SMBH becomes nearly coplanar
with the binary orbital plane on an another time scale $t_{\rm al1}
\sim 10^4 \, {\rm yr} \ga t_{\rm al2}$ only if the binary orbital
angular momentum $J_{\rm b} = a m v_{\rm b} \simeq m \left(GM
a\right)^{1/2}$ is larger than the rotation angular momentum $J_{\rm
BH} = a_* G M^2 /c$ of the central black hole. Taking $a\sim a_{\rm
m}$, we have 
\begin{equation}
{J_{\rm b} \over J_{\rm BH}} \simeq 6.8 \times 10^2 \left({a_* \over
            0.1}\right)^{-1} q_{-1}^{19/14} \alpha_{-2}^{2/7}
            \dot{m}_{-1}^{-3/14} M_8^{-3/7} ,
\label{eq:orbbhang}
\end{equation}
which gives $J_{\rm b} > J_{\rm BH}$ for
\begin{equation}
q \ga 8 \times 10^{-4} \left({a_* \over 0.1}\right)^{14/19}
              \alpha_{-2}^{-4/19} \dot{m}_{-1}^{3/19} M_8^{6/19} \sim
              q_{\rm cr} .
\label{eq:twq}
\end{equation}
Eq.~(\ref{eq:twq}) implies that {\it the spin axis of the central
SMBH formed during galaxy merging within Hubble time change its
orientation from vertical to the outer accretion disk to well aligned
with the rotation axis of the binary on a time scale $t \la 10^5 \,
{\rm yr}$}. As the inclination angle of binary plane is random, the
angle of the spin axis of the central SMBH relative to the galactic
plane distributes randomly.

When the secondary interacts with the accretion disk and distorts the
inner disk, it also twists and warps the accretion disk outside its 
orbit to some radius $r_{\rm al}$, as the quadrupole contribution to
the potential causes the precession of the major axis of an elliptical
orbit with frequency \citep{ivanov99}
\begin{equation}
  \Omega_{\rm ap} = {3 \over 4} q \left({a \over r}\right)^2
       \Omega_{\rm K} ,
\label{eq:quprec}
\end{equation}
where $\Omega_{\rm K}$ is the Keplerian angular velocity at $r$. The 
radius $r_{\rm al}$ out to which the disk is aligned with the orbital
plane depends on the quadrupole component of the binary 
gravitational field and the transfer of warp and angular momentum in 
the disk, and can be estimated simply, assuming that at the radius
$r_{\rm al}$ the time scale for radial transfer of the warp, $t_{\rm
warp}$, is on the order of the local quadrupole precession time scale
$\Omega_{\rm ap}^{-1}$. From $t_{\rm wp} = \Omega_{\rm ap}^{-1}$ and
Eq.~(\ref{eq:wq}), we have 
\begin{equation}
  r_{\rm al} \sim \left(q \alpha\right)^{1/2} \delta^{-1} a 
\label{eq:ral}
\end{equation}
for an accretion disk with $\alpha > \delta$ and diffusive-like
transfer of disk warp, or 
\begin{equation}
  r_{\rm al} \sim {3^{1/2} \over 2} q^{1/2} \delta^{-1/2} a 
\label{eq:ral2}
\end{equation}
for a disk with $\alpha < \delta$ and wave-like transfer of disk warp
of $t_{\rm wp} \sim r/c_{\rm s}$. The realignment radius $r_{\rm al}$
given by Eq.~(\ref{eq:ral}) or (\ref{eq:ral2}) is exactly the same as
the one given by \citet{ivanov99} and \citet{kumar90}. To have the
alignment scale larger than the binary orbit radius, we need
\begin{equation}
  q > 5\times 10^{-4} \alpha_{-2}^{-6/5} {\dot m}_{-1}^{2/5}
      M_8^{-1/5} x_3^{1/10} .
\label{eq:qal}
\end{equation}
Here we have used Eq.~(\ref{eq:delta1}) to obtain Eq.~(\ref{eq:qal}).
Eqs.~(\ref{eq:qcr}) and (\ref{eq:qal}) suggest that for any
SMBBHs-disk system formed during galaxy merging within Hubble time 
the secondary always aligns the accretion disk gets from the vicinity 
of the primary SMBH out to a radius $r_{\rm al} > a$. If accretion
disk is a slim disk with $\dot{m} \ga 1$ and $\delta \sim 1 > \alpha$,
Eq.~(\ref{eq:ral2}) implies $q \ga {4 \over 3} \delta \sim 1$ for 
$r_{\rm al} > a$. Therefore, only major merger can twist and align a
slim disk.

\subsection[]{Interaction between SMBBHs and ADAFs}

As for an ADAF the accretion gas has a quasi-spherical morphology with 
$\delta \sim 1$ and the orbital velocity $v_\phi$ of the gas is
significantly sub-Keplerian, the interaction between the secondary and
the accretion flow is insensitive to the inclination of the orbit
relative to the angular momentum vector of the accreting gas. 

As an ADAF has $\delta(\sim 1) > \alpha$, transfer of its any
distortion (warp and twist) is wave-like \citep{papaloizou85,musil02}.
As the accretion flow has a very low accretion rate and is 
quasi-spherical, its warp and twist
due to interaction with the secondary is negligible. The drag of the
accretion flow to the secondary SMBH can be approximated with the
interaction of a uniform gas of density \citep{narayan00}
\begin{equation}
  \rho = {\dot{M} \over 4 \pi r^2 v_{\rm r}} \simeq { \dot{m} 
    \dot{M}_{\rm Edd} \over 4 \pi r^2 \alpha_{\rm AD} V_{\rm K}} .
\label{eq:dens}
\end{equation}
As the moving SMBH has a relative velocity $|\vec{v}_{\rm rel}| \sim
v_{\rm K}$ to the gas, the drag force is \citep{ostriker99} 
\begin{equation}
  F_{\rm df} = - 4 \pi I \left({Gm \over v_{\rm rel}}\right)^2 \rho ,
\end{equation}
where the coefficient $I$ depends on the Mach number ${\it M} \equiv
v_{\rm rel} /c_{\rm s}$ and $c_{\rm s}$ is the sound velocity. 
As in our problem, $v_{\rm rel} \sim V_{\rm K} \ga c_{\rm s}$ with
${\it M} \ga 1$ and $I$ is approximately \citep{ostriker99,narayan00}
\begin{equation} 
  I \sim \ln(R_{\rm max} / R_{\rm min}) ,
\end{equation}
where $R_{\rm max} \sim H \sim r$ is the size of the system and
$R_{\rm min}$ is the effective size of the secondary, which is
approximately the accretion radius $R_{\rm min} \sim r_{\rm a} = 2 G m
/ v_{\rm rel}^2$. Therefore, we have 
\begin{equation} 
I \sim \ln \left({r v_{\rm rel}^2
  \over 2 G m}\right) = \ln\left({M \over 2 m}\right) \simeq 2 - \ln q_{-1} .
\end{equation}
Thus, the hydrodynamic drag time scale is 
\begin{eqnarray}
  t_{\rm hd} & \equiv & {m v_{\rm K} \over |F_{\rm df}|} \simeq {
    \alpha_{\rm AD} \over I \dot{m} } q^{-1} t_{\rm s} \nonumber \\
     & \simeq & 3\times 10^9 \, {\rm yr}\, q_{-1}^{-1}
    \left({\alpha_{\rm AD} \over 0.15}\right) \dot{m}_{-2}^{-1} , 
\end{eqnarray}
where $\dot{m}_{-2} = \dot{m}/10^{-2}$. Therefore, the hydrodynamic
drag of an ADAF on an orbiting SMBH is negligible.

When the secondary moves in an ADAF, some amount of gas is accreted
and another approximately the same amount of mass obtains velocities
greater than the escape velocity, as that in a standard thin disk (see
Sec~\ref{sec:stand}). Although it is different in ADAFs as the flow is
significantly sub-Keplerian and $\alpha_{\rm AD} \sim 0.3$ and
$\delta \sim 1$, Eq.~(\ref{eq:qcol}) can give a reasonable lower
estimate to the upper limit $q_0$. When $q < q_0$, the accretion flow
can compensate the out flow due to the collision of the secondary and
the accretion flow. Eq.~(\ref{eq:qcol}) gives $q_0 \sim \sqrt{3
\pi \over 8} \alpha^{1/2} \delta \simeq 0.6$. Therefore, the accretion
flow can compensate the mass outflow due to the SMBBHs-disk collision
even for a major merger.

\subsection[]{Realignment timescale of the orbital plane and the outer
  disk and the lifetime of AGNs}
\label{sec:outal}

When the twisted inner disk and the binary orbital plane 
become coplanar, the system stays quasi-stationary. As
it is determined by the specific angular momentum of the gas 
entering the disk, the orientation of outer accretion disk with $r >
r_{\rm al}$ is determined by the outer gas system and is supposed to
be nearly aligned with that of the galactic disk. The gas in the
accretion disk at $r > r_{\rm al}$ is accreting through the twisted
disk and exchanges angular momentum with the binary, leading to the
rotation axis of the aligned inner system slowly processing and
realigning with that of the outer disk plane \citep{rees78,ivanov99},
similar to the re-alignment of a rotating black hole with an inclined
accretion disk due to the Bardeen-Peterson effect \citep{scheuer96,
natarajan98}. The realignment time scale for a disk with $\alpha \sim
1$ is \citep{ivanov99}
\begin{equation}
  t_{\rm al3} \sim {J_{\rm b} \over \dot{J}_{\rm d}} \simeq {a m v_{\rm
      b} \over \dot{M} r_{\rm al} v_{\rm d}(r_{\rm al})} \simeq \left({a
      \over r_{\rm al}}\right)^{1/2} t_{\rm acc} < t_{\rm acc} 
\label{eq:tal3}
\end{equation}
where $\dot{J}_{\rm d}$ is the angular momentum flux of the disk at
$r_{\rm al}$ and $v_{\rm d}(r_{\rm al}) \simeq v_{\rm K}(r_{\rm al})$
is the disk angular velocity at $r_{\rm al}$. Defining a disk viscous
time scale  
\begin{equation}
  t_{\rm d} \equiv {M_{\rm d} \over \dot{M}} = \eta t_{\rm acc} ,
\label{eq:dvt}
\end{equation}
and from Eqs.~(\ref{eq:ral}), (\ref{eq:delta1}), and (\ref{eq:mmd}). we
have 
\begin{eqnarray}
  t_{\rm al3} & \sim & \left({a \over r_{\rm al}}\right)^{1/2}
              \eta^{-1} t_{\rm d} \nonumber \\
	      & \simeq & 0.60 q_{-1}^{3/4} \alpha^{1/2} 
	      \dot{m}_{-1}^{-1/2} M_8^{-0.61} x_4^{1/40} t_{\rm d} ,
\label{eq:tal3_1}
\end{eqnarray}
where $x_4 = r_{\rm al} / 10^4 r_{\rm G}$. 

However, for an accretion disk in AGNs, the viscous parameter $\alpha$
is $\ll 1$ and the case is more complex \citep{scheuer96,natarajan98}.
The effect of the binary on the accretion disk is to force the
rotation axis of the disk to process and to align with the binary
orbital plane and by Newtonian third-law the binary orbit realigns
with the accretion disk due to the feedback effect. Both precession
and alignment take place on the same time scale similar to the
realignment of rotating SMBH due to the Bardeen-Peterson effect, i.e. 
\begin{equation}
  t_{\rm al3} \sim \Omega_{\rm ap}^{-1} ,
\end{equation}
where $\Omega_{\rm ap}$ is the precessing angular velocity at $r_{\rm
al}$. From~(\ref{eq:quprec}), (\ref{eq:ral}), (\ref{eq:delta1}), and
(\ref{eq:dvt_1}), this gives
\begin{equation}
  t_{\rm al3} \sim 0.49 \alpha_{-1}^{29/10} q_{-1}^{3/4} \dot{m}_{-1}^{-3/10}
        M_8^{0.79} x_3^{53/40} t_{\rm d} ,
\label{eq:tal3_2}
\end{equation}
where $x_3 = a / 10^3 r_{\rm G}$. 
Eqs.~(\ref{eq:tal3_2}) and (\ref{eq:tal3_1}) give a similar result
that it takes about half of the disk viscous time to realign
the binary orbit with the outer accretion disk and  that $t_{\rm
al3} \sim 10^7 - 10^8 \, {\rm yr}$. As the interaction of binary-disk
takes place at $a > a_{\rm cr}$, Eq.~(\ref{eq:tal3}) suggests that the
binary orbital plane should become coplanar with the accretion disk 
before the separation $a$ becomes $\ll a_{\rm cr}$. Therefore, the spin
axis of the rotating central SMBH should be vertical to the accretion
disk when the secondary merges into the primary due to gravitation
wave radiation.  

From Eqs.~(\ref{eq:dvt}), (\ref{eq:tdm}) and (\ref{eq:dsize}), we have 
\begin{equation}
  t_{\rm d} = 2.2 \times 10^8 \, {\rm yr} \, \alpha_{-2}^{-4/5}
  \dot{m}_{-1}^{-2/5} M_8^{0.556} .
\label{eq:dvt_1}
\end{equation}
The lifetime of AGNs is a very important parameter in determining the
fueling mechanism of AGNs and the SMBH growth. The most estimates of
the net lifetime are in the range $t_{\rm Q} = 10^7 - 10^8 \, {\rm 
yr}$ for luminous Quasars with central SMBHs of mass $M \sim 10^8
- 10^9 {\rm M_\odot}$ \citep{haehnelt98a,martini01,steidel02,yu02a}
and most probably $t_{\rm Q} \simeq 5 \times 10^7 \, {\rm yr}$
\citep{jakobsen03} and $t_{\rm Q} \sim 6.6 \times 10^5 \, {\rm yr}$ for
miniquasars with central SMBHs of mass $M \sim 10^5 {\rm M_\odot}$ 
\citep{haiman98,haiman99} (for recent review see \citet{martini03}). It is
generally believed that the accretions in luminous QSOs and in
miniquasars are approximately at the Eddington rate. Taking the
fiducial value of the parameters, $\alpha = 0.03$, $\dot{m} = 1$ and
$M_8 = 1$ for luminous QSOs and $M_8 \sim 10^{-3}$ for miniquasars, we
have $t_{\rm d} = 3.6\times 10^7 \, {\rm yr}$ for luminous QSOs and
$t_{\rm d} = 7.8 \times 10^5 \, {\rm yr}$ for miniquasars which are
consistent with the estimates in the literature. Therefore, we take
the disk viscous time scale $t_{\rm d}$ 
as an indicator of the lifetimes of an accretion disk and of AGNs. 
 Eq.~(\ref{eq:dvt}) implies that SMBBHs in AGNs should merge 
within the viscous time scale of an accretion disk. While 
Eq.~(\ref{eq:tal3_2}) suggests that binary orbital plane and the
rotating central SMBH stays misaligned with the outer accretion disk
with a random inclination angle with respect to the outer accretion
disk for a great fraction of the viscous time of an accretion disk 
and of the life time of AGNs. 

When the binary orbital orbit becomes coplanar with the accretion disk, 
the secondary black hole opens a gap in the accretion disk and exchanges
angular momentum with outer disk gas via gravitational torques, leading 
to shrink of the binary separation on a viscous time scale $t_{\rm acc}$
\citep{lin86,ivanov99,armitage02}. The interaction of SMBBHs and a
coplanar accretion disk and the final coalescence of SMBBHs has been
discussed in detail by \citet{liu03}.


\section[]{Jet orientation and the formation of X-shaped feature
  in radio galaxies}
\label{sec:formation}

In Sec.~\ref{sec:model} and Sec.~\ref{sec:dbhit}, we discussed the
possible SMBBHs-disk system and their interaction in a sub-parsec 
scale. As relativistic jets in radio sources most likely initiate
forming before the two supermassive black holes become bound,
the relativistic jets may have developed to a large-scale size when
the binary becomes hard and interact with accretion disk.  It is 
believed that relativistic plasma jets form along the spin axis of
central SMBH and are perpendicular to the accretion disk due to
Bardeen-Peterson effect \citep{rees84,marscher02}, 

When the orbital radius of binary is less than the radius
$a_{\rm m}$ and the accretion disk is still a standard $\alpha$-disk
with accretion rate $\dot{m} >\dot{m}_{\rm cr}\sim 10^{-2} - 10^{-1}$,
the secondary realigns the central rotating SMBH via binary-disk
interaction, leading to reorientations of its spin axis and of the
relativistic jets with a reorientation time scale $t_{\rm al1} \la 10^5 
\, {\rm yr}$. As a relic radio lobe can be detected within
about $t_{\rm relic} \sim 10^6 - 10^8 \, {\rm yr}$ depending on the 
environment \citep{komissarov94,slee01,kaiser02}, the reorientation of
radio jets may be observed in some radio sources. Actually, the
observed X-shaped (or winged) radio sources \citep{leahy92,dennett02}
might be such objects.

\subsection[]{Summary of observations of X-shaped radio sources}

X-shaped, or winged, radio galaxies \citep{hogbom74} are a subclass of
extra-galactic radio sources of very peculiar morphology: a second
axis of symmetry of two large-scale old diffuse wings or tails,
orienting at an angle to the currently active lobes
\citep{dennett02,capetti02}. Many observations have been done to these
sources and show that (1) the winged sources are about 7 per cent of
the sample radio galaxies investigated by \citet{leahy92} and (2) they
are low-luminous FRII or borderline FRI/FRII radio galaxies and none
of them belongs to FRI-type \citep{dennett02}; (3) there is no 
evidence for a merger with a large galaxy in the last $\sim
10^8 \, {\rm yr}$ \citep{dennett02} in the sources except B2 0828+32
\citep{ulrich96} and 3C 293 \citep{evans99,martel99}; (4) all are
narrow-line galaxies except 4C +01.30 which show weak broad emission
lines \citep{wang03}; (5) the old wings as long as, or even longer
than, the directly powered active lobes have no pronounced spectral
gradients and form due to a jet reorientation within a few Myr
\citep{dennett02}; (6) the wings are aligned with 
the minor axis and vertical to the major axis of the host galaxy
\citep{capetti02,wang03}; (7) the active radio lobes have a
random inclination angle relative to the major axis of host galaxy;
(8) the host galaxy has a high eccentricity \citep{capetti02};
and (9) the wings are Z-symmetric \citep{gopal03}.

\subsection[]{Models in literature and their difficulties}
\label{sec:others}

Several scenarios have been suggested for the formation of the X-shaped
structure in some FRII radio sources in literature: (1) back-flow of
radio plasma from the active lobes into wings via buoyancy
\citep{leahy84} or diversion through the galactic disk
\citep{capetti02}; (2) slow conical precession of jet axis
\citep{parma85}; (3) quick reorientation of jet axis with or 
without turnoff of jet formation for some time \citep{dennett02,
merritt02}. \citet{dennett02} review all the models and find that the
first two are inconsistent with the observations and the third 
scenario together minor galaxy mergers is favored. 

A rapid reorientation of jet axis may result from the realignment of
a rotating SMBHs due to Bardeen-Petsreon effect with a misaligned
accretion disk which forms due to the disk instabilities
\citep{dennett02}, or from the spin-flip of the active SMBH due to the
coalescence of an inclined binary black holes \citep{dennett02,
merritt02,zier01}. However, disk instabilities would be suppressed 
even by a mild rotation of the SMBH and cannot explain the 
straightness of jet from VLBI- to VLA-scale \citep{pringle96}. 
Another difficulty with the disk instability model is in explaining 
why such instability does not exit in other radio galaxies which have
stable jet direction and why it occurred only once in the X-shaped 
sources \citep{dennett02}. 

\citet{merritt02} suggest that the rapid change of jet orientation 
may be due to a spin-flip of the central active black hole due to a
coalescence of misaligned SMBBHs. To explain the detection rate of
X-shaped radio sources, they show that the merger has to be minor. 
Although the observations of X-shaped radio sources favor minor merger
scenario \citep{dennett02,gopal03}, the spin-flip picture has several
defects. First of all, inclined rotating SMBH formed via the binary
coalescence should realign with the accretion disk due to the
Bardeen-Peterson effect on a short time scale $t_{\rm al1} \la 10^5 \,
{\rm yr}$, implying that the relativistic jets reorients in the
direction of the old wings on the time scale and we should detect a
distorted jet of a length $l_{\rm j} \sim t_{\rm al1} v_{\rm j} \la 10
\, {\rm kpc}$ for a typical jet velocity $v_{\rm j} \sim 0.3 c$
instead of straight wings of hundreds Kpc. Secondly, Merritt \& Ekers
do the calculation with Newtonian approximation, while calculations 
basing on general relativity show that the change in inclination of 
a rotating central SMBH is negligible in a minor merger and a
significant reorientation of the active SMBH requires a comparatively
rare major merger \citep{hughes03}. Thirdly, as we show in
Sec.~\ref{sec:dbhit} that binary-disk interactions would align a
central SMBH with an inclined binary orbital plane before the binary 
coalesces and that no change in the orientation of spin axis of the
central SMBH happens even in a major merger. The last is 
from the observations. The model does not reasonably explain the sharp
transition that X-shaped feature is detected only in FRII radio 
galaxies but not in FRI radio galaxy with similar luminosity.

\subsection[]{Reorientation of radio jets in FRII radio galaxies}
\label{sec:jor}

Here we suggest that the formation of X-shaped feature in some radio
sources is due to the realignment of the rotating central SMBHs with
the binary orbital plane via binary-disk interaction and the
Bardeen-Peterson effect. In this scenario, the accretion disk and the 
gas in galactic disk have already settled down and large scale
relativistic jets in radio galaxies have formed, before the secondary
SMBH distorts the accretion disk. It is expected that the accretion
disk is nearly coplanar with the dust lane or galactic disk due to the
conservation of angular momentum of gas. Thus, the large scale
relativistic radio jets and lobes are nearly vertical to the galactic
plane. 

If the accretion rate is greater than the critical rate $\dot{m}_{\rm
cr} \sim 10^{-2} - 10^{-1}$ and the accretion disk is a standard 
thin-disk, the twisted disk reorients the spin axis of the rotating 
SMBH, leading to the rapid change of jet direction on a time scale
$t_{\rm reor} \sim t_{\rm al1} \la 10^5 \, {\rm yr}$. The relic of the
old radio lobes forms the detected old wings in the X-shaped radio
sources. As the jet before the reorientation is vertical to the 
accretion disk, the wings are expected to be nearly perpendicular to
the galactic plane, as is observed. 

The winged radio source 4C +01.30 shows weak broad emission lines and
contains a partially obscured quasar nucleus. The mass of its central
SMBH is $M \sim 4\times 10^8 \, {\rm M}_\odot$ and the accretion disk
is a standard $\alpha$-disk with $\dot{m} = L / L_{\rm Edd} \simeq 0.2$ 
\citep{wang03}. From Eq.~(\ref{eq:at1}), the reorientation takes place
on a time scale $t_{\rm reor} \simeq 2 \times 10^5 \, {\rm yr}$ for
typical parameters $a_* = 0.7$,  $\alpha = 0.03$ and $\epsilon =
0.3$. From Eq.~(\ref{eq:alt2}), the disk becomes twisted and warped
due to the interaction with the secondary and coplanar with the binary
orbital plane on a time scale $t_{\rm al2} \simeq 4 \times 10^4 \, {\rm
yr}$. From Eq.~(\ref{eq:dvt_1}), the viscous time of the sources is
$t_{\rm life} \simeq 1.5\times 10^8 {\rm yr}$. The mass ratio of
accretion disk and the secondary SMBH is $ \eta \approx 7 q_{-1}^{-1}$
and $\eta > 1$ even for a major merger with $q \sim 0.7$. 

As the other X-shaped radio galaxies show only narrow emission lines,
the central quasar nuclei may have been completely obscured by the
dust torus and the X-shaped sources are edge-on. It is easy to be 
understood as observations prefer to detecting X-shaped feature in
radio sources with large scale projected jets. Therefore, the high
eccentricity of the host galaxy of X-shaped radio sources is most
likely due to selection effects and does not relate to the origin of
the X-shaped structure. This might be the reason why in the control
sample of radio galaxies used by \citet{capetti02} the radio galaxies
with host galaxy of a similar or even higher eccentricity do not show
winged feature.

\subsection[]{The missing of winged FRI radio galaxies}

It is possible that the accretion rate has become less than 
the critical rate $\dot{m}_{\rm cr}$ and accretion disk is no
longer a standard $\alpha$-disk but a radiatively inefficient
accretion flow, e.g. an ADAF, when the interaction between the 
secondary and an accretion disk starts. As the interaction between 
an ADAF and a binary is negligible, reorientation of the spin
axis of the central SMBH takes place on a very long time scale $t_{\rm  
hd} \sim 10^9 \, {\rm yr}$ which is longer than the observable
time scale of a radio relic $t_{\rm relic} \sim 10^7 - 10^8 \, {\rm
  yr}$ (see the discussion below). As the transition of FRI and FRII
radio galaxies is related to an accretion rate $\dot{m}_{\rm FR} \sim
\dot{m}_{\rm cr}$ \citep{ghisellini01,cavaliere02,maraschi03}, the FR
division may reflect the transition of accretion mode from a standard
disk to a radiatively inefficient accretion flow and the accretion 
disk in FRI radio galaxies is radiatively inefficient. This
implies that the binary-disk interaction cannot form a winged
structure in FRI radio sources and the missing of X-shaped FRI radio 
sources is due to the radiatively inefficient accretion mode, i.e. 
ADAF, ADIO, or CADF. However, it is possible that the outer part of 
accretion disk is standard and the inner accretion flow is an ADAF. 
In this case, the interaction between the binary and the outer 
accretion disk can twist the part of standard disk on a time scale
$t_{\rm al2} \sim 10^3 \, {\rm yr}$ but the inner ADAF realign the
central rotating black hole on a time scale $t_{\rm al1} \sim 10^7 \,
{\rm yr}$ for $\alpha_{\rm AD} \sim 0.15$ and $\dot{m} \sim 10^{-2}$
which is on the same order of the relic time $t_{\rm relic}$. Here, we
use Eq.~(\ref{eq:at1}) to estimate the realignment time scale for a
transition system $ \dot{m} \dot{m}_{\rm cr}$. Therefore,
it is expected that S-shaped structure would be observed in 
high-luminous FRI radio sources.

One possibility to detect X-shaped feature in FRI radio sources is 
that radio sources evolves from FRII type into FRI type after the
wings forms with $t_{\rm al3} > t_{\rm d}$. From
Eqs.~(\ref{eq:tal3_2}), this implies  
\begin{equation}
  q > 1.5 \left({\alpha \over 0.05}\right)^{-58/15} \dot{m}_{-2}^{2/5}
  M_8^{-1.05} x_3^{-53/30} .
\label{eq:xfri}
\end{equation}
As the mass ratio $q$ should be smaller than unit, Eq.~(\ref{eq:xfri})
implies that no X-shaped FRI radio galaxy is possible to form.

\subsection[]{Evidence against recent mergers}

The undisturbed properties of host galaxies of winged radio sources
except 3C293 imply that mergers are minor or, if major, longer than a
few $10^8 \, {\rm yr}$ ago. Eq.~(\ref{eq:tbd}) shows that two SMBHs
become bound on a time scale $\sim 10^8 \, {\rm yr}$. A bound binary
becomes hard and interacts with an accretion disk at a pc-scale on a
similar time scale (cf \citet{yu02}). Therefore, the galaxy merging in
X-shaped radio sources may occur $\sim 10^9 \, {\rm yr}$ ago in our
scenario which is consistent with the observations of 3C 293 
\citep{evans99}. 

3C293 is the only winged source showing obvious signs of interaction
of a tidal tail and a close companion galaxy \citep{evans99,
martel99}. The relative masses of its host galaxy and the companion  
suggest that the tidal feature is most likely a remnant from a merger
event occurring more than $t_{\rm td} \sim 10^9 \, {\rm yr}$ ago
\citep{evans99}. The spectroscopy observations of central bulge region 
show strong CO emission line with a velocity width $\sim 400 \, {\rm
km \; s^{-1}}$. From Eq.~(\ref{eq:mass}), the mass of the central SMBH
is $M \simeq 2.2 \times 10^9 \, {\rm M_\odot}$. If we take $t_{\rm td}
\sim 10^9 \, {\rm yr} \sim 2 t_{\rm bd}$, Eq.~(\ref{eq:tbd}) gives a
mass ratio $q \sim 0.3$, implying that the merger in 3C293 is 
minor but with a moderate mass ratio and is
rare. This is consistent the observations that the merger occurred
quite long time ago but still can be detected and that 3C293 is the
only X-shaped radio source showing signs of interaction. The mass 
ratio of the disk and the secondary is $\eta \sim 9 \alpha_{-2}^{-4/5}
\dot{m}_{-1}^{3/5}$ and $\eta > 1$ for $\dot{m} > \dot{m}_{\rm
  cr}$. Therefore, the binary black hole will become realigned with
the outer accretion disk and merger into one more massive SMBH,
leading to the formation of a double-double FRII radio galaxy. From
Eq.~(\ref{eq:am}, the interaction happens likely at $\la 3.6\times
10^2 r_{\rm G} \alpha_{-2}^{4/7} \dot{m}_{-1}^{-3/7} \sim 0.08 \, {\rm
pc}$, which is much less than the disk size $r_{\rm d} \simeq 0.4 \,
{\rm pc}$.

\subsection[]{Detection rate of winged radio sources and the lifetime
  of low luminosity FRII radio galaxies}
\label{sec:age}

\citet{leahy92} show that the probability $\lambda_{\rm X}$ of
detecting a FRII radio source with X-shaped radio feature is
$\approx 7 \%$ in a sample of radio galaxies with luminosity between
$3 \times 10^{24}$ and $3 \times 10^{26} \, {\rm W\; Hz^{-1}}$ at 1.4
GHz. The detection rate depends both on the mean observable time scale,
$t_{\rm relic}$, of relic radio lobes and the minimum $t_{\rm min}$
between the life time $t_{\rm life}$ of FRII radio galaxies and the
mean detectable 
time $t_{\rm merger}$ between mergers, $t_{\rm min} = \min(t_{\rm 
life},t_{\rm merge})$. It is difficult to accurately estimate the mean
timescale $t_{\rm relic}$ in a survey sample. The estimated time scale
$t_{\rm relic}$ is in the range of $\sim 10^6 - 10^8 \, {\rm yr}$,
depending on both the environment and the survey frequency
\citep{komissarov94,slee01,kaiser02}, which is consistent with the
spectrally estimated age limit of radio wings in some X-shaped
sources: $< 34 \, {\rm Myr}$ for 3C223.1 and $< 17 \, {\rm Myr}$ for
3C403 \citep{dennett02}, and $< 75 \, {\rm Myr}$ for B2 0828+32
\citep{klein95}. 

From the measured rate and the timescale $t_{\rm relic}$, we have
\begin{equation}
  t_{\rm min} = {t_{\rm relic} \over \lambda_{\rm X}} \la 10^9 \, {\rm
  yr} \, \left({ t_{\rm relic} \over 10^8 \, {\rm yr}}\right) .
\label{eq:xdet}
\end{equation}
\citet{merritt02} take the upper limit
$t_{\rm relic} \sim 10^8 \, {\rm yr}$ and get $t_{\rm min} \simeq 
10^9 \, {\rm yr}$, which is too large to be the age of radio sources. 
They interpret $t_{\rm min}$ as the mean merge time scale of galaxy,
$t_{\rm merge}$, which is higher than the estimate of the galaxy 
merge rate given in literature \citep{haehnelt98,carlberg00} although
it is not implausible. However, a relic with an age $t_{\rm relic} 
\sim 10^8 \, {\rm yr}$ is most likely invisible even at a very low
survey frequency due to expansion and radiation. To use the detection
rate given by \citet{leahy92} for a sample of low luminous FRII radio
galaxies with a high survey frequency 1.4 GHz, it is most plausible to
adopt $t_{\rm relic} \approx 10^7 \, 
{\rm yr}$, which gives $t_{\rm min} \simeq 10^8 \, {\rm yr}$. As the 
life time $t_{\rm life}$ of low luminosity FRII radio galaxies is much
less than the mean merge time scale $t_{\rm merge}$ of galaxy which is
$\ga 10^9 \, {\rm yr}$, we take $t_{\rm min}$ as the mean lifetime of
low luminous FRII radio sources and have $t_{\rm life} \sim 10^8 \,
{\rm yr}$, which is consistent with the estimated lifetime $t_{\rm
life} \sim 10^8 \, {\rm yr}$ for low-luminous AGNs in literature. As
in low luminous FRII radio galaxies the accretion rate is $\dot{m}
\ga \dot{m}_{\rm FR}$, we take a mean accretion rate $\dot{m} \sim 
0.1 $ for the Leahy \& Parma's sample and from Eq.~(\ref{eq:dvt_1})
obtain a disk viscous time $t_{\rm d} \simeq 10^8 \, {\rm yr }$ for
$\alpha = 0.03$. The estimates of the lifetime of low luminous FRII
radio sources in different ways are consistent with each other very
well.

\subsection[]{Orientations of wings and active radio lobes in FRII 
  radio sources}
\label{sec:jetfrii}

As an accretion disk forms with gas of low angular momentum in a
merging system of galaxies and the gas is settled down into the
galactic plane with low gravitational potential. It is expected that 
large-scale relativistic plasma jets are perpendicular to accretion
disk due to the Bardeen-Peterson effect and to the galactic plane.
Since the radio wings in X-shaped radio sources are the relics of
radio jets and lobes, they should be vertical to the major axis of 
host galaxy. While the orientations of active jets are aligned with
the rotation axis of binary and distribute randomly. This is 
consistent with the observations of X-shaped radio galaxies 
\citep{capetti02} that the wings in all the winged radio galaxies are
nearly aligned with the minor axis of host galaxy and the active lobes
have no preferential orientation. 

To reorient the rotating central SMBH, the accretion disk has to be
standard with $\dot{m} > \dot{m}_{\rm cr}$ and thus
the radio galaxies morphologically belong to FRII-type. However, if 
binary hardening time scale ($\sim t_{\rm bd}$) is much larger
than the disk viscous time $t_{\rm d}$, the accretion disk 
becomes an radiatively inefficient disk, e.g. ADAF, with
$\dot{m} < \dot{m}_{\rm cr} \sim \dot{m}_{\rm FR}$ before binary-disk
interaction. Thus, a radio source evolves from FRII- into FRI- class
without the formation of X-shaped radio structure. 
Eqs.~(\ref{eq:tbd}) and (\ref{eq:dvt_1}) suggest that 
X-shaped radio structure cannot form in a galaxy merging with 
a mass ratio 
\begin{equation}
  q \ll q_{\rm X} \equiv 3 \times 10^{-2} M_8^{-0.08}
  \alpha_{-2}^{1.07} \dot{m}_{-2}^{0.54} .
\label{eq:qx}
\end{equation}
Therefore, the mass ratio in any X-shaped FRII radio galaxies should
be $q \ga 10^{-2}$, which is consistent with the estimates of
the X-shaped sources 3C293 and 4C+01.30. From Eqs.~(\ref{eq:orbbhang}) 
and (\ref{eq:qx}), we have $J_{\rm b} \gg J_{\rm BH}$ even for a
central SMBH with $a_* \sim 1$. Eqs.~(\ref{eq:qcr}) and (\ref{eq:qx})
imply that nearly all of the hard binary systems formed during galaxy
merging within Hubble time would produce X-shaped feature. 

When a binary twists inner accretion disk and aligns the central SMBH,
the orbital plane stays inclined for a time scale $t_{\rm al3}$ which 
is a great large fraction of the disk viscous time $t_{\rm d}$. It is
expected that large scale jets in most of FRII radio sources would 
randomly orients with respect to the major axis of host galaxy. 
The inclined radio jets in radio galaxies are nearly aligned from the
VLBI to the VLA as $t_{\rm al3} \sim 10^7 - 10^8 \, {\rm yr} > t_{\rm
relic} \ga 10^7 \, {\rm yr}$. When the binary orbital plane becomes
coplanar with the outer accretion disk, radio jet becomes vertical 
to the galactic plane and aligned with the minor axis of host galaxy.
After the binary becomes coplanar with outer accretion disk, they
merge on a short time scale and the radio source becomes a DDRG
\citep{liu03}. Therefore, the fraction of FRII radio galaxies with
vertical large scale radio jets is small. When a radio source becomes
a DDRG, the mass in the inner accretion disk has been removed,
accelerating the evolution of the disk. It is expected that the
accretion disk becomes radiatively inefficient on relatively short
time $< t_{\rm d}$. Therefore, the fraction of FRII radio galaxies
with random orientation of jet is determined by the ratio $t_{\rm
al3}/t_{\rm acc} \sim 1$. 

The random orientations of jets in FRII radio galaxies may have been
observed. It is found that the jet orientations
in radio sources randomly distribute relative to the dust lane or the
major axis of host galaxy \citep{birkinshaw85,schmitt02}. There is no
significant correlations between the misalignment angle and any of the
intrinsic kinematic parameters of host galaxy, in particular rotation
velocity and central velocity dispersion which is related to the mass
of central SMBH. \citet{schmitt02} show that the dust disks are
closely aligned with the major axis of host galaxy and the jets are
well aligned from the VLBI to the VLA scales. None of the possible
mechanisms for the origin of the observed misalignment between jet and
the rotation axis of host galaxy could consistently explain the
observations \citep{schmitt02}. In our scenario, jet orientation is 
determined by the impact angle of the merging galaxy and the
misaligned angle between radio jet and the minor axis in FRII radio 
galaxies should be independent of any intrinsic kinematic parameter of
the parent galaxy. However, the model suggests that the misalignment
occurs in an accretion disk at pc scale inside the broad 
line region (BLR). As the emission line flux depends on the ionization
radiation which is a function of the inclination angle between the
inner radiation region and BLR and the BLR cover factor. A positive 
correlation between the misalignment angle and the relative line flux
may be expected. 

Two more important implications of the model are that the distribution
of jet orientation depends on the FR type and that the jets in DDRGs
are nearly vertical to the major axis of host galaxy. We will discuss 
the two predications in more detail in Sec.~\ref{sec:jetfri} and
Sec.~\ref{sec:ddrg}.

\subsection[]{Relationship of FRI and FRII radio sources and
  jet orientations in FRI radio sources}
\label{sec:jetfri}

As all young AGNs have been detected as FRII radio galaxies, it is
most likely that FRI radio sources is evolved from FRII-type 
\citep{odea98}. 
There are three possible ways for a radio source to evolve from FRII- 
to FRI-type in our scenario. If the activity of a FRI radio source
is triggered by minor mergers with mass ratio $q \ll q_{\rm X}$, the
evolution finishes before the binary-disk interaction. In those 
sources (Class I FRI radio sources), the change of jet orientation due 
to the interaction may occur on a time scale $t_{\rm al1} \sim 10^7 -
10^8 \, {\rm yr}$, leading to the formation of S-shaped structure in
FRI radio sources. As the realignment time scale $t_{\rm al1}$
inversely correlates with the accretion rate $\dot{m}$, S-shaped radio
structure are most likely to form in luminous FRI radio sources with 
accretion rate close to $\dot{m}_{\rm FR} \sim \dot{m}_{\rm cr}$. The
jet orientations in most the ClassI FRI radio sources are random
relative to the galactic plane of host galaxy. Since ClassI FRI radio
sources should contain binary of mass ratio $q_{\rm cr} \sim
5 \times 10^{-4} \ll q \ll q_{\rm X} \sim 10^{-2}$, they should
make up $\la 1/5$ of FRI radio sources. Class FRI radio galaxies have
linear size larger than their progenitor FRII radio galaxies do.

While most radio sources with larger mass ratio $q \ga q_{\rm X}$ 
spend much more time on FRII phase, which have more energetic 
radio jets with larger size. The second possible case is that the
alignment time scale $t_{\rm al3}$ in a FRII radio source is much
larger than the disk viscous time scale $t_{\rm d}$ and the accretion
disk becomes radiatively inefficient before the outer disk-binary
realignment. The jets in these subclass FRI radio sources (Class II)
randomly orients. ClassII FRI radio sources have an average linear 
size larger than that of ClassI FRI radio galaxies. From 
Eq.~(\ref{eq:tal3_2}), the activity of ClassII FRI radio sources 
must be triggered by major mergers with $q \gg 0.1$. Therefore, 
the ClassII FRI radio galaxies with random jet orientation and largest
size make up a very small fraction of FRI radio sources as major
mergers in galaxy merging is very rare. 

Most radio sources evolve from FRII- into FRI-type after
the binary plane becomes coplanar with outer accretion disk and 
the galactic plane. \citet{liu03} and the discussion in
Secs.~\ref{sec:jetfrii} and \ref{sec:ddrg} show that SMBBHs become
merged on a time scale $\la t_{\rm acc}$ and FRII radio galaxies
become DDRGs, before evolving from FRII- to FRI-type. Radio
jets in this subclass of FRI radio sources (Class III) should be
vertical to the galactic plane and aligned with the minor axis of host
galaxy. After the galaxies become DDRGs and the radio jets restart,
the accretion rate becomes $1 \gg \dot{m} \ga \dot{m}_{\rm cr} \sim
\dot{m}_{\rm FR}$. The re-born radio sources have FRII
morphology with jet less powerful than that in normal FRII radio
galaxies. Since the formation of jet is interrupted for quite long
time before the radio sources evolves fro FRII- to FRI-types, the
size of active radio lobes in ClassIII FRI radio sources is the size
of the re-born sources and as large as that of the ClassI FRI radio
sources which is much smaller than the size of most FRII radio
galaxies. Therefore, our conclusion is that the average linear size
of FRI radio galaxies is smaller than that of FRII radio galaxies. 
However, the relics of up to four giant radio lobes could be detected
in some ClassIII FRI radio galaxies as the sources become giant
when they become DDRGs \citep{liu03}.

The two predications about jet orientations and source size of 
radio sources can be tested. We note that \citet{dekoff00} suggests
that the jet orientations in FRII radio galaxies randomly distributes
while the jets in most FRI radio sources are vertical to the major
axis of host galaxy. However, the conclusion is based on a small 
sample of radio galaxies and should be checked with much larger sample
of radio sources.

\section[]{Relationship between X-shaped and double-double radio
  galaxies} 
\label{sec:ddrg}

In a minor mergers of mass ratio $q \ga q_{\rm X}$, the binary orbital
plane and the accretion disk becomes coplanar on a time scale $t_{\rm
  al3} \sim 10^7 - 10^8 \, {\rm yr}$ after an X-shaped radio structure
forms in a FRII radio galaxy. The secondary SMBH opens a gap in the 
accretion disk and exchanges angular momentum with disc gas via
gravitational torques for a minor merger with $q > q_{\rm min} = (81
\pi/ 8) \alpha \delta^2 \simeq 3 \times 10^{-5} \alpha_{-2} \delta_{-2}^2$
\citep{lin86}. The secondary migrates inwards on a viscous timescale
$\sim t_{\rm acc}$ and merges into the primary, leading to the removal
of the inner accretion disk and to an interruption of jet formation
\citep{liu03}. \citet{liu03} identify DDRGs \citep{schoenmakers00a} 
with the objects in which coalescence of SMBBHs, removal of inner 
accretion disk and interruption of jet formation occurred. 
DDRGs are a subclass of giant FRII radio galaxies, 
consisting of a pair of symmetric double-lobed structures with one
common center. The new-born inner structure with
relative low luminosity is well aligned with the outer old lobes. The 
generation of inner lobes in DDRGs is due to the interaction of warm 
clouds and recurrent jets of interruption time $\sim \, {\rm My}$
\citep{kaiser00,schoenmakers00a}. When interruption time is $\ll 10^6
{\rm yr}$, the interaction of the recurrent jet and intergalactic
medium could not produce new lobes \citep{kaiser00}, as may be the
case in the non-DDRG recurrent sources 3C288 \citep{bridle89}, 3C219
\citep{clarke92}, and B1144+352 \citep{schoenmakers99}.  

The scenario of the binary coalescence and disk removal \citep{liu03} 
implies that nearly all the SMBBHs in FRII radio galaxies merge into
one more massive black hole and that all the winged radio sources
would evolve into double-double radio galaxies on a time scale
$t_{\rm X-DD} \sim t_{\rm acc}$. This picture is consistent with that 
double-double lobes are detected only in FRII radio galaxies. 
As the outer accretion disk is nearly aligned with the outer
galactic plane, it is expected that the restarting jets in DDRGs
is vertical to the major axis of host galaxy and aligned with 
the old radio wings in its X-shaped progenitor. We are statistically 
testing this predication and our preliminary results confirm the
predication that the jets in DDRGs are well aligned with the minor
axis of host galaxy \citep{liu03a}.

As the active jets in DDRGs is aligned with the old wings and $t_{\rm
X-DD} \sim 10^7 - 10^8 \, {\rm yr} \ga t_{\rm relic}$, it is
impossible to detect the co-existence of the wings and the
double-double lobes in one DDRG. However, it is possible to detect the
coexistence of double-double radio lobes and the cavities in the IGM,
excavated by the past plasma jet with randomly orientation and filled
with back-flow radio plasma from the outer double lobes. The cavities
formed in such a way should have straight and sharp edges toward the
core on the opposite side and diffusive edges on the same side of the
active radio jets. The observations of the FRII radio galaxy J0116-473
may fit the description. \citet{saripalli02} show that J0116-473 is a
low-luminosity FRII radio galaxy and contains both double-double radio
lobes and a bar-like feature with sharply bounded northern edge. The
observations show that the bar-like feature may have an age of $\sim
10^8 \, {\rm yr}$ and the present activity restarts about $\sim (3-4) 
\times 10^6 \, {\rm yr}$ ago. The elapsed time since the last energy
supply to the outer giant lobes is smaller than $< 7\times 10^7 \,
{\rm yr}$ \citep{saripalli02}. If J0116-473 contained SMBBHs once
before and the bar-like structure is the cavity produced by the past
misaligned jet and filled with the back-flow plasma due to the
alignment of the binary orbital plane with outer accretion disk 
on a time scale $t_{\rm al3} \sim t_{\rm acc}$. The estimated age
$t_{\rm acc} \sim  10^8 \, {\rm yr}$, of the bar-like feature together
with Eq.~(\ref{eq:acc}) implies that the mass ratio $q \simeq 0.2
\dot{m}_{-1}$ and the merger is minor with $q > q_{\rm X}$.

Since DDRGs evolve from X-shaped radio galaxies, it is expected that
double-double radio lobes like X-shaped feature should be detected
only in FRII radio sources. The probability $\lambda_{\rm DD}$
to detect recurrent jets in FRII radio sources is $\lambda_{\rm DD}
\sim t_{\rm DD} / t_{\rm life}$, where $t_{\rm DD} = t_{\rm int} + 
t_{\rm tr}$ with $t_{\rm int}$ the interruption time scale of jet 
formation and $t_{\rm tr} = l_{\rm j} / v_{\rm lobe}$ is the time
scale for relativistic plasma lobes to travel along jet from central
engine to outer relic radio lobes. For a typical value of advancing 
velocity $v_{\rm lobe} \sim 0.2 c$ and a typical length scale of giant
radio sources $l_{\rm j} \sim 1 \, {\rm Mpc}$, we have $t_{\rm tr} \sim
10^7 {\rm yr}$. As the interruption time scale $t_{\rm int} \sim \,
{\rm Myr}$, we adopt $t_{\rm DD} \simeq 10^7 \, {\rm yr}$ and obtain 
$\lambda_{\rm DD} \la 5$ per cent. From Eq.~(\ref{eq:xdet}), we have 
\begin{equation}
  \lambda_{\rm DD} = \left({t_{\rm DD} \over t_{\rm relic}}\right)
          \lambda_{\rm X} < \lambda_{\rm X} 
\label{eq:dddet}
\end{equation}
with $t_{\rm DD} < T_{\rm relic}$. Eq.~(\ref{eq:dddet}) suggests that
the detection rate of DDRGs in a sample of FRII radio sources is
between $ (1 - 10) \%$. In estimating the time scale $t_{\rm tr}$, we
used the length scale of giant radio sources as \citet{liu03} suggest
that the radio galaxies would become giant when they become DDRGs. As
the accretion rate to produce the restarting relativistic jets is much
smaller than the one producing the relic outer lobes, the active jets
in DDRGs are much less luminous and may not be able to reach the 
outer lobes on the time scale $\sim t_{\rm relic}$. If so, $t_{\rm tr}
\sim t_{\rm relic}$ and $\lambda_{\rm DD} \simeq \lambda_{\rm X}$. If
the giant radio galaxies form mainly due to the explosive increase of
accretion rate via the interaction of SMBBHs and accretion disk,
the detection rate of recurrent jets in a sample of giant radio 
galaxies may be as high as $\sim 100 \%$.


\section[]{Discussions and conclusions}
\label{sec:dis}

Galactic dynamical simulations show that SMBBHs may stall, when it
becomes hard at $a_{\rm h} \sim 0.01 - 1 \, {\rm pc}$, and merge on a
timescale may longer than Hubble time. We show that the interaction of
accretion disk and SMBBHs may dominate the evolution of a binary
formed during galaxy merging within Hubble time, after
it becomes hard. We investigate the interaction of inclined SMBBHs and
an accretion disk at sub-parsec scale and its feasibility to
significantly change the orientation of the spin axis of central SMBH,
which is believed to be aligned with the orientation of relativistic
jets in AGNs. We identify the interaction and the reorientations of
spin axis of central black hole with jet orientations in some radio 
sources. 

It is shown that the inclined secondary twists the accretion disk and
aligns the inner accretion disk. We analytically calculate the
alignment time scale and show that the alignment finishes on a time
scale $\sim 10^3 \, {\rm yr}$ which is the same result as the one
obtained with numerical computations by \citet{ivanov99} who also
numerically calculate the alignment. We show that the quick alignment
of the inner accretion disk is slowed down at the Bardeen-Peterson
radius $r_{\rm BP} \sim 20 r_{\rm G}$ and completed on an another
timescale $t_{\rm al1} \sim  10^4 \, {\rm yr}$ due to the 
Bardeen-Peterson effect, which is associated with the reorientation
timescale of relativistic plasma jets. We suggest that the
reorientation of spin axis of central rotating SMBH causes the
formation of X-shaped feature observed in some FRII radio galaxies. In
our scenario, the wing of X-shaped radio sources is the relic of past
jets and lobes and its alignment with the minor axis of host galaxy is
due to the conservation of angular momentum of accreting gas and the
alignment of the accretion disk and the galactic plane. The random
distribution of orientation of active jet is due to the random impact
of two merging galaxies. The back-flow-diversion model 
\citep{capetti02} suggests that a back-flow is driven out along the
minor axis of host galaxy due to the largest pressure gradient in the
direction. However, the picture cannot explain why many radio galaxies
with similar luminosity and eccentricity of host galaxy have no
X-shaped feature and why the size of wings in some X-shaped radio
sources are much larger than the directly powered lobes. Although the 
spin-flip model has theoretical problem if a merger is minor, it is
possible to have old wings along the minor of host galaxy if the 
merger is major and the coalescence of two black holes finishes on 
a time scale $\ll t_{\rm al1} \sim 10^4 \, {\rm yr}$. The difficulty
with the scenario is that the new-born jet is distorted due to the
Bardeen-Peterson effect and becomes realigned with the old wings on a
time scale $\la 10^5 \, {\rm yr}$, which implies a distorted jet
of size $\la 10 \, {\rm Kpc}$ instead of a straight large scale
radio jets. 

To be more specification, we consistently explain the observations of 
X-shaped radio galaxies with our scenario and in particularly apply it
to the two winged sources 4C+01.30 and 3C293. We show that the mass
ratio of the secondary and the primary in the 4C+01.30 and 3C293 are
indeed minor with  $q \sim 0.2$ and the reorientation happens on time
$\sim 2\times 10^5 \, {\rm yr}$ which is consistent with the
observation of $\la {\rm Myr}$. 

Basing the model and the detection rate $\sim 7 \% $ of X-shaped
structure in a low luminous FRII radio galaxies \citep{leahy92}, 
we estimate the average lifetime of low luminous FRII radio sources 
to be $t_{\rm life} \sim 10^8 \, {\rm yr}$ if taking a mean observable
timescale, $t_{\rm relic} \sim 10^7 \, {\rm yr}$, of the relic of
radio lobes in Leahy \& Parma's sample. The estimate of the lifetime
of low luminous FRII radio sources is consistent with the estimates of
low luminous AGNs and the disk viscous time scale but much larger
than the estimate for QSOs which may accrete at the Eddington
accretion rate. This is reasonable if the activity of AGNs is
triggered by a the minor merger. The theoretical calculation also
shows that the interaction of accretion disk and a binary formed by a
minor merger can lead to the formation of X-shaped structure in radio
galaxies. In the back-flow model, wings form by the back-flow plasma in
each radio galaxies with high eccentricity \citep{capetti02}, which
suggests the observation of X-shaped structure in each radio galaxy
with similar eccentricity of host galaxy and is inconsistent with the
observations.  \citet{merritt02} take $t_{\rm relic} \simeq 10^8 \,
{\rm yr}$  in the spin-flip model and obtain $t_{\rm life} \sim 10^9
\, {\rm yr}$. They suggest that the time scale $t_{\rm life}$ is the
mean merger time scale of galaxy and obtain a coalescence rate of
SMBHs $\sim 1 \, {\rm Gyr^{-1}}$ which is consistent with those
inffered for galaxies in dense regions or groups but higher than most
estimates of the overall galaxy merger rate \citep{haehnelt98,
carlberg00}. However, the
explanation has two difficulties. The first of is that the relic of
radio lobes cannot be detectable, especially at a survey frequency as
high as 1.4GHz, on a time scale $t_{\rm relic} \sim 10^8 \, {\rm yr}$
due to radiation loss and plasma expansion. The other is that the mean
life time of radio sources is much shorter than the mean merger
timescale and the estimate $t_{\rm life}$ should be the average
lifetime of radio sources with low luminosity between $3 \times
10^{24} \, {\rm W\; Hz^{-1}}$ and $3 \times 10^{26} \, {\rm W\;
Hz^{-1}}$ at 1.4GHz. 

In our model, the lack of X-shaped FRI radio galaxies may be due to
the fact that the accretion flow in FRI radio galaxies is
geometrically thick and optically thin and weakly interacts with
SMBBHs. Or it may also be due to that FRI radio galaxies are evolved
from FRII radio galaxies on a time scale much longer than the
detectable time scale of relic lobes $t_{\rm relic}$. 
As the accretion disk in FRII is standard with high accretion rate
$\dot{m} > \dot{m}_{\rm cr}$ while radiatively inefficient accretion
flow with accretion rate $\dot{m} < \dot{m}_{\rm cr}$, it is most
likely that radio galaxies evolves from FRII- to FRI-type. We
discussed the three kinds of possible evolutions and suggest that 
the orientation of jet in most but not all FRII radio galaxies 
distributes randomly while is nearly vertical to galactic plane in
most but not all FRI radio galaxies. The different distribution of jet
orientation in different FR-type radio galaxies may be observed 
\citep{dekoff00}. \citet{schmitt02} discussed all the possible
explanations in literature to the detected distribution of jet
orientation and concluded that none of them is plausible. Here we give
a reasonable model. However, the sample is too small to be very
meaningful and the result needs to be confirmed with larger samples of
radio galaxies.  

Our model also suggests that all SMBBHs in FRII radio
galaxies become coplanar with galactic plane and merge into a more
massive black hole within the life time of the FRII radio galaxies.
\citet{liu03} suggest that the coalescence of SMBBHs in FRII radio
galaxies leads to the formation of DDRGs. This implies that X-shaped
FRII radio galaxies forms ahead of DDRGs on a viscous time scale
$t_{\rm acc} \sim 10^8 \, {\rm yr}$. We applied the configuration to
the DDRG J0116-473 which also shows a bar-like feature much older than
the relic outer lobes. If we suggest that the bar-like feature in
J0116-473 form due to the refill with back-flow radio plasma of the
cavities in IGM excavated by the past plasma jet, the observations
imply that the merger is minor with a mass ratio $q \sim 0.1$.

We divided FRI radio sources into three subclasses according to
the different relation to FRII radio galaxies. ClassI FRI radio
sources are evolved from FRII radio sources before the interaction of
accretion disk and SMMBHs. As in this subclass the stage of FRII phase
is shorter than the average life time of most FRII radio galaxies, it
may be expected that the average size is shorter than that of FRII
radio sources. While ClassIII FRI radio sources form after SMBBHs get
merged and relativistic jets restarts with lower power, the average
size of the subclass is determined by the recurrent jets and is also
expected to be smaller than that of FRII radio sources. One 
expectation is that it is possible to detect relic of outer giant
radio lobes in some FRI radio sources, which should be aligned with the
inner active radio lobes. If $t_{\rm acc} \ll t_{\rm relic}$, one may
even detect a third pair of outer lobes with oldest age and S-shaped
structure. From Eq.~(\ref{eq:acc}), this implies that $q \ll 10^{-2}
\dot{m}_{-2} \left(t_{\rm relic} / 10^7 \, {\rm yr}\right)$. Such triple
pairs of radio lobes may have been detected in the nearest AGN Cen A
(NGC5128) \citep{israel98}. However, to compare the observation of
CenA and the model in more detail, we need the estimates of ages of
the lobes. 

Our model implies that ClassI and ClassII FRI radio galaxies and FRII
radio galaxies with random jet orientation with respect to the
galactic plane harbor SMBBHs with separation of $10^2 r_{\rm G} \la a
\la 10^3 r_{\rm G}$. A close binary of $a < a_{\rm cr} \sim 10^2
r_{\rm G}$ is a source of gravitational radiation. The coalescence of
SMBBHs can produce an enormous burst of gravitational radiation. As
the orbital plane of binary of $a < a_{\rm cr} \sim 10^2 r_{\rm G}$ is
coplanar with the accretion disk and the system is old, it may be
expected to detect in low luminous FRII radio sources with jets nearly
vertical to the galactic disk close binary, which gives rise strong
gravitational wave radiation and would be good targets for the
monitoring of gravitational wave interferometers, e.g. LISA. Although
the FRII radio galaxies in which binary merged also shows vertical
radio jets, it is easy to distinguish them from the former subclass of
FRII radio galaxies as they may have passed through the DDRG phase and
contain giant relic of outer radio lobes.

No X-shaped structure is observed in high luminosity FRII radio
galaxies and QSOs, which may be due to selection effect 
\citep{dennett02}. The accretion may be at the Eddington accretion
rate $\dot{m} \sim 1$ and the accretion disk is slim but not
standard in QSOs. From Eqs.~(\ref{eq:tbd}) and (\ref{eq:dvt_1}), we
have $t_{\rm bd} \simeq 14 t_{\rm d} \alpha_{-1}^{4/5} \dot{m}^{2/5}
M_8^{0.06} q_{-1}^{-3/4.02} \gg 1$, which implies that the missing of 
high luminous FRII radio galaxies and QSOs may be due to no
binary-disk interaction during the phase of high accretion rate. In
the QSO, it is expected that the jets are nearly aligned with the
minor axis of host galaxy. If the binary interacts with accretion disk
during the phase of QSO and luminous radio galaxies, the transfer of
warp is in a wave-like way and the time scales $t_{\rm al1}$ and $t_{\rm
al2}$ are smaller than with those given by Eq.~(\ref{eq:at1}) and 
(\ref{eq:alt2}), leading to a rapid reorientation of jet in QSO. 
Since the time scale $t_{\rm al3} \sim 4 \times 10^6 \, {\rm yr}$ 
is too short for the jet to form a large scale X-shaped structure, we
could detect distorted radio jet in QSOs and luminous radio galaxies.

\section*{Acknowledgments}

We thank Dr. Z.H. Fan for helpful comments and Prof. D.N.C. Lin and
Prof. X.-B. Wu for pleasant discussions. This work is supported by the 
National Natural Science Fundation of China (NSFC 10203001) and 
the National Key Project on Fundamental Researches (TG 1999075403).

\label{lastpage}

\end{document}